\documentclass[twocolumn,floatfix,groupedaddress,superscriptaddress,aps,prb,showpacs]{revtex4-1}

\usepackage{graphicx}
\usepackage{amsmath}
\usepackage{amssymb}
\usepackage{amsfonts}
\usepackage{dsfont}
\usepackage{dcolumn}
\usepackage{bbm}

\usepackage{color}
\begin{document}

\title{Conservation laws protect dynamic spin correlations from decay:\\
Limited role of integrability in the central spin model}

\author{G\"otz S.\ Uhrig}
\email{goetz.uhrig@tu-dortmund.de}
\affiliation{Lehrstuhl f\"{u}r Theoretische Physik I, 
Technische Universit\"{a}t Dortmund,
 Otto-Hahn Stra\ss{}e 4, 44221 Dortmund, Germany}

\author{Johannes Hackmann}
%\email{johannes.hackmann@tu-dortmund.de}
\affiliation{Lehrstuhl f\"{u}r Theoretische Physik II, 
Technische Universit\"{a}t Dortmund,
 Otto-Hahn Stra\ss{}e 4, 44221 Dortmund, Germany}

\author{Daniel Stanek}
%\email{daniel.stanek@tu-dortmund.de}
\affiliation{Lehrstuhl f\"{u}r Theoretische Physik I, 
Technische Universit\"{a}t Dortmund,
 Otto-Hahn Stra\ss{}e 4, 44221 Dortmund, Germany}
 
 \author{Joachim Stolze}
%\email{joachim.stolze@tu-dortmund.de}
\affiliation{Lehrstuhl f\"{u}r Theoretische Physik I, 
Technische Universit\"{a}t Dortmund,
 Otto-Hahn Stra\ss{}e 4, 44221 Dortmund, Germany}
 
 \author{Frithjof B. Anders}
%\email{frithjof.anders@tu-dortmund.de}
\affiliation{Lehrstuhl f\"{u}r Theoretische Physik II, 
Technische Universit\"{a}t Dortmund,
 Otto-Hahn Stra\ss{}e 4, 44221 Dortmund, Germany}

\date{\textrm{\today}}

\begin{abstract} 
%%Key I
Mazur's inequality renders statements about persistent correlations possible. 
We generalize it in 
a convenient form applicable to any set of linearly independent constants of motion.
%%Key II
This approach is used to show rigorously that a fraction of the initial spin  correlations
 persists indefinitely in the isotropic central spin model unless the average
coupling vanishes.
The central spin model
describes a major mechanism of decoherence in a large class of
potential realizations of quantum bits. Thus the derived results contribute significantly
to the understanding of the preservation of coherence.
%%Key III
We will show that persisting quantum correlations are not linked to 
the integrability of the model, but caused by a finite operator overlap
with a finite set of constants of motion.
\end{abstract}

%78.67.Hc 	Quantum dots
%78.67.Bf 	Nanocrystals, nanoparticles, and nanoclusters
%72.25.Rb 	Spin relaxation and scattering
%03.65.Yz 	Decoherence; open systems; quantum statistical methods (see also 03.67.Pp in quantum information; for decoherence in Bose-Einstein condensates, see 03.75.Gg)
%02.30.Ik 	Integrable systems

\maketitle

\paragraph{Introduction.}
The two-time correlation function of two observables reveals
important information about the dynamics of a system in and out of
equilibrium: The noise spectra are obtained from symmetric
combinations of correlation functions, while the causal, antisymmetric
combination determines the susceptibilities required for the theory of
linear response.

The two-time correlation function only depends on the time
  difference if at $t=0$ the system of interest is prepared in a  stationary state
  whose density operator commutes with the time-independent Hamiltonian. 
  This is what will be considered in this work.
Since correlations generically decay
for $t\to\infty$, important information about the system dynamics is
gained if a non-decaying fraction of correlations prevails at infinite
times.
Such non-decaying correlations are clearly connected to a limited
dynamics in certain subspaces of the Hilbert space. The question
arises if such a restricted dynamics is always linked to the
integrability of the Hamiltonian.
Here integrability means that the Hamiltonian 
can be diagonalized by Bethe ansatz which implies that there is
an extensive number of constants of motion.
Identifying and understanding those non-decaying correlations can
be potentially exploited in applications for persistent storage of
(quantum) information.

In this Letter we first prove that persisting correlations are
not restricted to integrable systems by using a generalized form
of Mazur's inequality \cite{mazur69,suzuk71}. This is in contrast to
the behavior of the Drude weight in the 
frequency-dependent conductivity of one-dimensional systems which appears
to vanish abruptly once the integrability is lost, even if only by including
an arbitrarily small perturbation. So far, the Drude weight has been
the most common application of  Mazur's inequality, see for instance 
Refs.~\onlinecite{zotos97,jung07,sirke11,prose11a} and references therein. 
Second, we apply
this approach to the central spin model (CSM) \cite{Gaudin1976,*gaudi83}
describing the interaction of a single spin, e.g., an electronic
spin in a quantum dot \cite{schli03,hanso07}, an effective two-level
model in a NV center in diamond \cite{jelez06}, or a $^{13}$C nuclear
spin \cite{alvar10c}, coupled to a bath of surrounding nuclear spins
inducing decoherence.

Persisting spin correlations have been found in the
CSM  by averaging the central spin dynamics over a bath of random
classical spins \cite{merku02,chen07} or in Markov approximation
\cite{khaet03,coish04}. Finite-size calculations
\cite{stane13a,hackm14a} of the full quantum problem and stochastic
evaluation \cite{farib13a,*farib13b} of the exact Bethe ansatz
equations \cite{Gaudin1976,*gaudi83} for small system sizes $(N\le48)$ have also provided
evidence for a non-decaying fraction of the central spin correlation,
predicting a non-universal, system dependent value.  Its origin has
remained obscure, and it has been speculated that the lack of spin
decay might be linked to Bose-Einstein condensate-like physics
\cite{farib13a}.

While it is fascinating to identify such non-decaying correlations, it
is technically very difficult to rigorously establish them.
Approximate methods often miss precisely those intricate aspects
allowing correlations to persist, especially when they explicitly
exploit the assumption that the system relaxes towards a statistical
mixture. Numerical approaches are either restricted in system size
\cite{dobro03a,hackm14a}, or they are limited in the maximum
time which can reliably be captured \cite{gober05,stane13a,hackm14a}.
Even analytical solutions \cite{Gaudin1976,*gaudi83} can often only be
evaluated in small systems \cite{farib13a,*farib13b}.
Thus, a rigorous result establishing the existence of non-decaying
correlations is highly desirable and we resort to Mazur's inequality for this purpose.

\paragraph{General Derivation.}
To establish the key idea and to fix the notation we 
present the following modified derivation related to Suzuki's 
derivation in Ref.\ \onlinecite{suzuk71}.
We consider the time-indepen\-dent Hamiltonian $H$ and the
operator $A$ with a vanishing expectation value $\langle
A\rangle=0$ with respect to a stationary density operator $\rho$,
i.e.\ $[\rho,H]=0$ so that two-time correlation functions only depend on 
the time difference. Note that $\rho$ does not need to be the equilibrium density operator.  
Then, $\rho$ and $H$ have a complete
common eigenbasis $\{ |j\rangle\}$ in a finite-dimensional Hilbert space, and
their spectra are $\{ \rho_j >0 \} $ and $\{ E_j\} $, respectively. 
We define the correlation function  of  $A$  as 
\begin{subequations}
\begin{eqnarray}
\label{eq:s-def}
S(t) &:=& \langle A^\dagger(t) A(0)\rangle
= \mbox{Tr}\left[\rho A^\dagger(t) A(0)\right]
\\
\label{eq:lehmann}
&=& \sum_{j,m} \rho_j |A_{jm}|^2 \exp(i(E_j-E_m)t),
\end{eqnarray}
\end{subequations}
so that Eq.\ \eqref{eq:lehmann} is its Lehmann representation,
and $A_{jm}:=\langle j|A|m\rangle$ denotes
the matrix element of $A$. Physically, $S(t)$ stands for 
a measurement of $A^\dagger$ at time $t$ after the evolution
from the initial state prepared by applying $A$ at $t=0$. Especially, for $A=S^z$
of a spin $S=1/2$ in a disordered environment, $S(t)$ is proportional to
 $\langle S^z(t)\rangle$ if $\langle S^z(0)\rangle=1/2$,
see Supplement A for details.
If $\lim_{t\to\infty} S(t)$ exists, it  is given by 
\begin{equation}
\label{eq:s-infty}
S_\infty := \sum_{jm} \rho_j |A_{jm}|^2\delta_{E_j,E_m} \ge 0
\, .
\end{equation}
If $S(t\to\infty)$ does not exist, and
$|S(t)|<\infty$, the long-time average
$\lim_{T\to\infty} T^{-1}\int_0^TS(t)dt=S_\infty$ is projecting out
the time-independent part $S_\infty$ and uniquely defines  the
non-decaying fraction of the correlation.  
 
In practice, the Lehmann representation \eqref{eq:lehmann} requires 
the complete diagonalization of $H$ which 
is not feasible for large systems. Hence one resorts to constants of motion.
To this end, we
define the scalar product for two operators $X$ and $Y$ as
\begin{equation}
\label{eq:scalar}
(X|Y):=\langle X^\dagger Y\rangle = \mbox{Tr}\left[\rho X^\dagger Y\right]
\end{equation}
in the super-Hilbert space of the operators.
If a set of $M$ conserved linearly independent
operators $X_i$ with $[X_i,H]=0$ is known, one may
assume their orthonormality $(X_i|X_m)=\delta_{im}$ provided by a 
Gram-Schmidt process.
Then, we expand the operator of interest $A$
\begin{eqnarray}
\label{eq:operator-expansion}
A &=& \sum_{i=1}^M a_i X_i  + R
\end{eqnarray}
in this incomplete operator basis where $a_i :=(X_i|A)$ and 
$R$ is the remaining rest with $(X_i|R)=0$ $\forall i\in\{1,\ldots,M\}$.
Substituting \eqref{eq:operator-expansion} into the definition \eqref{eq:s-def}
yields
\begin{equation}
\label{eq:pythagoras}
S(t) = \sum_{i=1}^M |a_i|^2 + S^{(R)}(t)
\end{equation}
with $S^{(R)}(t):=\langle R(t) R(0)\rangle$.
This relies on the constancy of (i) $\langle X_i^\dagger(t) X_m(0)\rangle=\delta_{im}$,
of (ii) $\langle X_i^\dagger(t) R(0)\rangle=0$, and of (iii) $\langle R^\dagger(t) X_m(0)\rangle=0$ all
stemming from $[X_j,H]=0$. 
For the last relation we have used the cyclic invariance of the trace
and $[\rho,H ]=0$.

If we knew $\lim_{t\to\infty}S^{(R)}(t)=0$, we would deduce 
$S_\infty = \sum_{i=1}^M |a_i|^2$.
But in general this does not hold because $R$ may still contain a non-decaying part.
But \eqref{eq:pythagoras} implies Mazur's inequality
\begin{equation}
\label{eq:lower_bound}
S_\infty \ge S_\text{low} := \sum_{i=1}^M |a_i|^2.
\end{equation}
For a given $H$, the complete set of conserved operators $\Gamma$ is spanned by all 
pairs of energy-degenerate eigenstates
\begin{equation}
\Gamma :=\left\{ {|j\rangle\langle m|}/{\sqrt{\rho_m}} \ \text{with} \ E_j=E_m \right\}.
\end{equation}
The elements of $\Gamma$ are orthonormal with respect to
the scalar product \eqref{eq:scalar}. The coefficent $a_{jm}$ of $X_{jm}={|j\rangle\langle m|}/{\sqrt{\rho_m}}$ 
\footnote{For clarity, we use a double index here.}
takes the value $\sqrt{\rho_m}A_{jm}$ so that the right hand side of \eqref{eq:lower_bound}
equals $S_\infty$ as given by the Lehmann representation \eqref{eq:s-infty}. Thus, the inequality
\eqref{eq:lower_bound} is tight because it becomes exact for the \emph{complete} set $\Gamma$
of conserved operators. The physical interpretation of Eq.\ \eqref{eq:lower_bound} is
straightforward in the Heisenberg picture if we view the time-dependent observable $A^\dag$ as super vector.
Its components parallel to conserved quantities (super vector directions) are constant in time because 
these quantities commute with the Hamiltonian. But all other components, which are perpendicular to the
conserved super subspace, finally decay.

If not all conserved operators are considered, the r.h.s.\ of
\eqref{eq:lower_bound} decreases and only the inequality
holds. Generally, if \emph{any} subspace of the space spanned by
$\Gamma$ is considered Mazur's inequality \eqref{eq:lower_bound}
holds. One does not need to know the complete set of
eigenstates of $H$ in order to calculate a lower bound: Any
finite (sub)set of conserved operators is sufficient.  

Now we proceed to generalize Mazur's inequality for easy-to-use application.
Usually, some conserved operators $C_i$ are known but they are
not necessarily orthonormal in general.  Rather their overlaps yield
a Hermitian, positive norm matrix $\mathbf{N}$ with matrix elements
$N_{im}:=(C_i|C_m)$.  Each operator $C_i$ can be represented as a
linear superposition of the complete set of orthonormal $X_i$.
These superpositions can be summarized in a matrix $\mathbf{M}$ so
that $\mathbf{c} = \mathbf{M}^* \mathbf{x}$ where the vectors
$\mathbf{x}$ and $\mathbf{c}$ contain the operators $X_i$ and $C_i$ as
coefficients; $\mathbf{M}^*$ is the complex (not Hermitian!)
conjugate of $\mathbf{M}$.  A short calculation shows that
$\mathbf{N}=\mathbf{M} \mathbf{M}^\dagger$.

If we define the vector $\mathbf{a}_X$ with complex components 
$a_i$, the  bound $S_\text{low}$ 
can be expressed by $S_\text{low}=\mathbf{a}_X^\dagger\mathbf{a}_X$.
In analogy, we compute $\mathbf{a}_C$ with complex components $(C_i|A)$. Obviously, 
$\mathbf{a}_X= \mathbf{M}^{-1} \mathbf{a}_C$ holds and the lower bound
is computed by
\begin{equation}
S_\text{low} = \mathbf{a}_C^\dagger (\mathbf{M}^{-1})^\dagger \mathbf{M}^{-1}\mathbf{a}_C
=\mathbf{a}_C^\dagger \mathbf{N}^{-1}\mathbf{a}_C
\label{eq:gen-low-bo}
\end{equation}
without resorting to orthonormalized operators, relying only
on the scalar products of $C_i$ and $A$. 
We have successfully eliminated the construction of a subset of orthogonal
operators $X_i$ and related the lower bound to some known set of 
linear independent unnormalized conserved operators $C_i$.
The general lower bound \eqref{eq:gen-low-bo}
is our first key result. A possible route to generalizations to various
initial states is sketched in the Supplement.

\paragraph{Central spin model.}
The Hamiltonian of the CSM  reads
\begin{equation}
\label{eq:hamilton}
H_0=\vec{S}_0\cdot\sum_{k=1}^N J_k \vec{S}_k
\end{equation}
where we assume all spins to be $S=1/2$ for simplicity. It is a
generic model to study the interaction between a two-level
system and a bath of spins or more generally a set of subsystems
with finite number of levels. Currently, it is intensively
investigated for understanding the decoherence and
dephasing in possible realizations of quantum bits
\cite{schli03,lee05,hanso07,petro08}.
Theoretical tools comprise Chebyshev polynomial technique
\cite{dobro03b,hackm14a}, perturbative approaches
\cite{khaet02,coish04,coish10}, generalized Master equations
\cite{fisch07,ferra08,barne12}, equations of motion \cite{deng08}
various cluster expansions \cite{witze05a,yang08a,maze08,cywin10},
Bethe ansatz \cite{Gaudin1976,*gaudi83,bortz07b,bortz10b,farib13a,farib13b},
density-matrix renormalization \cite{stane13a}, and studies of the
classical analogue \cite{merku02,erlin04,alhas06,chen07}.

By focusing on $A=S^z_0$, the correlation
function defined in \eqref{eq:s-def} reveals important information on
the decay of the central spin. Due to isotropy no other components
of the central spin need to be considered. Given the smallness of the hyperfine
couplings ($J_k$ is in the range of $\mu$eV corresponding to percents
of a Kelvin \cite{merku02,schli03,lee05,hanso07,petro08}) the
experimentally relevant temperature can be considered as infinite, and
we take the spin system to be completely disordered, i.e.,
$\rho\propto \mathds{1}$, prior to the preparation of an initial state
of the central spin, cf.\ Supplement. 

For classical spins $S_k$, there are strong analytical
arguments that a fraction of central spin correlations persists
unless there is a diverging number of arbitrarily weakly coupled spins
in the bath \cite{merku02,erlin04,chen07}. In the quantum case smaller
systems have been studied and evidence for a non-decaying
fraction of spin polarization \cite{farib13a,farib13b,hackm14a} has
only be compiled in fairly small ($N<50$) systems or up to fairly
short times \cite{stane13a}.

Based on the generalized Mazur's inequality \eqref{eq:gen-low-bo}, we are able to
address the nature and the lower bound of these non-decaying
correlations for arbitrary system sizes.
The total spin $\vec{I}:=\sum_{k=0}^N \vec{S}_k$ could serve as a
first guess for a useful conserved quantity.  Only the $z$-component
$C_1:=I^z$ has an overlap $\mathbf{a}=(I^z|S^z_0)=1/4$ (we omit the
subscript $_C$ for brevity).  The norm $N_{11}=(I^z|I^z)$ takes the
value $(N+1)/4$ so that \eqref{eq:gen-low-bo} provides
$S_\text{low}=1/(4(N+1))$. Irrespective of the considered distribution
of the couplings $J_k$, using only $I^z$ as single conserved
operator does not provide a meaningful lower bound for
thermodynamically large, or infinite baths.

The next important conserved quantity is the energy $H_0$ itself. But,
of course, $(H_0|S^z_0)=0$ because $H_0$ is a scalar and $S^z_0$ a
vector component.  The $z$-component of the product $\vec{I}H_0$,
$H^z_0:= I^z H_0$, clearly fulfills $[H_0,H^z_0]=0$ and defines
a conserved composite vector operator. We find
\begin{subequations}
\begin{eqnarray}
(S_0^z|H^z_0) &=& J_S/16
\\
(H^z_0|H^z_0) &=&(2J_S^2+3(N-1)J_Q^2)/64
\end{eqnarray}
\end{subequations}
where $J_S:=\sum_{k=1}^N J_k$ and $J^{2}_Q:=\sum_{k=1}^N J_k^2$. 
With this input Eq.\ \eqref{eq:gen-low-bo}
yields
\begin{equation}
\label{eq:ss-res1}
S_\text{low}=\frac{1}{4}\frac{J_S^2}{2J_S^2+3(N-1)J_Q^2}.
\end{equation}
This bound remains finite for $N\to\infty$ if the $J_k$ are drawn from
a probability distribution $p(J)$ with average $\overline{J}$ and
variance $\overline{\Delta J^2}$.  For large $N$ one has
$J_S=N\overline{J}$ and $J^2_Q= N(\overline{J}^2+\overline{\Delta
J^2})$ so that
$S_\text{low}={\overline{J}^2}/[20\overline{J}^2+12\overline{\Delta
J^2}]$ ensues for $N\to\infty$. This is a finite lower bound unless
the average values $\overline{J}$ vanishes.
This rigorous bound is our second key result.
%%%%%%%%%%%%%%%%%%%%%%%%%%%%%%%%%%%%%
\begin{figure}[tb]
\begin{center}
	\includegraphics[width=1.0\columnwidth,clip]{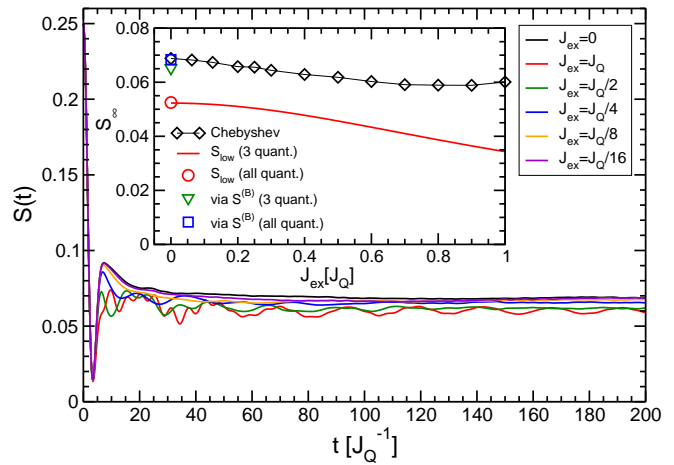}
\end{center}
\caption{(Color online) Spin correlation $S(t)$ for $N=20$ bath spins with $J_k\propto k$,
but normalized such that $J_Q=1$ is the unit of energy,
and various $J_\text{ex}$ defined in \eqref{eq:jex}. The inset compares $S_\infty$ from the average of the 
numerical data with $t\in[150/J_Q,200/J_Q]$ to  $S_\text{low}$ obtained
from \eqref{eq:gen-low-bo} for the 3 quantities ($I^z, I^z_Q, H_0^z$) or for all
 quantities ($I^z, H_l^z$  with $l\in\{1,2,\ldots N\}$). The
estimates from the Overhauser correlations $S^{(B)}$ are also shown.
\label{fig:ss}
}
\end{figure}
%%%%%%%%%%%%%%%%%%%%%%%%%%%%%%%%%%%%%%%%%%%%%%%%%%%%%%%%%%%%%%%%%%%%%%%%%%%%%%%

For any finite system with non-vanishing sum $J_S$, Eq.\
\eqref{eq:ss-res1} provides a rigorous finite lower bound which is
very easy to compute for any given set of couplings.  It can
serve to check the validity of numerical results such as provided in
Refs.\ \cite{farib13a,farib13b,hackm14a}.
Generically, distributions of the $J_k$ have finite
values $\overline{J}$ and $\overline{J^2}$. 
This is the case for nuclear spins in molecules \cite{alvar10c} or NV centers in diamond
\cite{jelez06} because the spin baths are finite.
In quantum dots, the convergence and existence
$J_S$ and $J_Q$ is ensured even for arbitrary number of spins
because the couplings are bounded from above, but become arbitrarily small due to
exponential tails of the electron wave function \cite{merku02,schli03,lee05,hanso07,petro08}.
This leads to vanishing $\overline{J}$ implying complete decay for infinite times.

For large, but finite times, however, our results include the possibility of slow decays $S(t)\propto
\ln(t)^{-\alpha}$ previously advocated for infinitely large spin baths
\cite{erlin04,chen07,li12}.  Assuming exponential scaling for the couplings
 $J_k \propto \exp(-\beta k)$, where $\beta$ is inversely proportional to 
 the number of relevant bath spins \footnote{In a quantum dot, this is the number of nuclear
 spins within the localization volume of the electronic wave function, typically 
 $10^{4}$ to $10^{6}$. It is not the total number of nuclear spins which is of the order
 of Avogadro's constant.}, it is clear that $J_S$ and $J^2_Q$ converge quickly for
$N\to\infty$ so that Eq.\ \eqref{eq:ss-res1} implies
$S_\text{low}\propto 1/N$.
Chen et al.\ \cite{chen07} have argued that at any given \emph{finite}
time $t$, only those spins $\vec{S}_k$ with couplings $tJ_k \gtrapprox
1$ significantly influence the real-time dynamics of the central
spin. Hence, only an effective number $N_\text{eff}(t)\propto \ln(t)$
of spins contribute to the correlation function implying $S(t)\propto
1/\ln(t)$ for such a distribution function.

The lower bound \eqref{eq:ss-res1} can be improved by considering the
three conserved observables $I^z$, $H_0^z$, and
$I^z_Q:=I^z\sum_{i<j}\vec{S}_i\vec{S}_j$.
The required vector and matrix elements are given in the supplemental
material.  Still the bound does not exhaust the numerically found
value as depicted in the inset of Fig.\ \ref{fig:ss} for
$J_\text{ex}=0$ ($J_\text{ex}$ makes the system non-integrable, 
it will be defined in \eqref{eq:jex}).
Even resorting to the integrability of the CSM
\cite{Gaudin1976,*gaudi83} which implies $0=[H_l,H_p]$ with $H_l:=\sum_{k=0,\ne
l}^N (\varepsilon_l-\varepsilon_k)^{-1} \vec{S}_l\cdot\vec{S}_k$ and
$\varepsilon_0=0, \varepsilon_k=-1/J_k$ does not account
for the full non-decaying fraction obtained in finite size
calculations \cite{hackm14a}, see circle in the inset of Fig.\
\ref{fig:ss}. The bound has been computed considering $I^z$ and
$H^z_l:=I^z H_l$ for $l\in\{1,2,\ldots N\}$ (for matrix elements see
supplement). 

The above results suggest that the integrability is not the key
ingredient for a finite non-decaying fraction.  To support this
claim we extend the Hamiltonian \eqref{eq:hamilton} by adding one
extra coupling $H_0\to H$
\begin{equation}
\label{eq:jex}
H:=H_0+J_\text{ex}\vec{S}_1\cdot\vec{S}_N
\end{equation}
between the most weakly and the most strongly coupled bath spin,
defined to be at $k=1$ and $N$, respectively. Its value $J_\text{ex}$
is chosen to be ${\cal O}(J_Q)$ so that it constitutes a sizable perturbation
even for large spin baths.

The
modified time-dependence of $S(t)$ is depicted for various
$J_\text{ex}$ in Fig.\ \ref{fig:ss}. A finite $J_\text{ex}$ spoils
the integrability completely \cite{Gaudin1976,gaudi83}, 
but leaves the quantities $I^z, I^z_Q, H^z$
conserved.  These three constants of motion generic for isotropic spin
models are used to obtain the lower bound (red curve) in
the inset of Fig.\ \ref{fig:ss}. Obviously, $S_\text{low}$ is
decreased smoothly and only moderately upon increasing $J_\text{ex}$
in line with the numerically determined $S_\infty$.  
There is no abrupt jump to zero, in contrast to what is known for the Drude weight.
The conclusion
that integrability is only secondary for the non-decaying spin
correlation is our third key result.

At present it remains an open question which conserved quantities one has to include to
yield a tight lower bound. We presume that higher powers of $H$, for instance $I^z H^2$, 
have to be considered. Such studies are more tedious and left for future research.
Instead, we take a mathematically less rigorous route based on the estimate by Merkulov et al.\
\cite{merku02}
\begin{equation}
\label{eq:estimate}
S_\infty=S^{(B)}_\infty/(12S^{(B)}(0))
\end{equation}
 where $S^{(B)}(t)$ is the correlation
of the Overhauser field operator
$\vec{B}_{\text{N}}:= \sum_{j=0}^N J_k \vec{S}_k$. Note that an arbitrary
$J_0$ can be included 
because $\vec{S}_0\cdot\vec{B}_{\text{N}}$ differs from $H_0$ in \eqref{eq:hamilton} 
only by an irrelevant constant for spin $1/2$. 
This estimate was derived for a classical, large Overhauser field \cite{merku02}
and prevails in the thermodynamic limit of  the quantum case:
The Overhauser field becomes a classical variable upon $N\to\infty$ as shown
in Ref.\ \cite{stane13a}.

Thus we now apply the general approach \eqref{eq:gen-low-bo} to 
$A=B^z_{\text{N}}$. Considering only $C_1=I^z$ as conserved operator already
yields a meaningful lower bound for the Overhauser field correlation function
for $N\to\infty$ 
\begin{equation}
\frac{S^{(B)}_\text{low}}{S^{(B)}(0)}
=\frac{(J_S+J_0)^2}{(N+1)(J_Q^2+J_0^2)}  \, .
\end{equation}
Recall $J_S\propto N$ and $J_Q^2\propto N$ if the couplings are drawn from a normalized
 distribution function $p(J)$.
This lower bound can be optimized by choosing the arbitrary value
$J_0$ such that the bound becomes maximal. With the matrix elements
given in the supplement $S^{(B)}_\text{low}$ can be improved
considering the three constants $I^z, I^z_Q, H^z$ or all integrals
$I^z$ and $H^z_l, 1\le l\le N$. The results are also included in
Fig.\ \ref{fig:ss} (triangle and square symbols). 
They hold only for $J_\text{ex}=0$ because the
estimate \eqref{eq:estimate} applies only in this case. Remarkably,
the resulting estimates for $S_\infty$ seem to be tight.  In
particular, the easily evaluated estimate based on all integrals
reproduces the numerically found $S_\infty$ to its accuracy.  We
applied the same estimate to the case $J_k \propto \exp(-\beta k)$ studied by
stochastically evaluating the Bethe ansatz equations and
found excellent agreement with the published data with $N\le48$ in Ref.\
\cite{farib13b} as well. Thus we conjecture that the non-decaying
fraction $S_\infty$ in the central spin model is quantitatively
described by $S^{(B)}_\text{low}/(12S^{(B)}(0))$ if
$S^{(B)}_\text{low}$ is determined from the $N+1$ integrals $I^z$ and
$H^z_l$. This constitutes our fourth key result. The small difference,
however, between triangle (from three constants of motion) and square 
(from $N+1$ constants of motion) in Fig.\ \ref{fig:ss} indicates
again that the significance of the integrability is limited.

In summary, four key results are obtained: (i) An easy-to-use version of Mazur's 
inequality to prove persisting correlations; (ii) A rigorous finite lower bound for
the infinite-time spin correlation in the CSM,
valid for the infinite system if the average coupling is finite; (iii)
Only a small part of the persisting correlation is due to the
integrability; (iv) A quantitative estimate for the persisting
correlation is conjectured, based on the Overhauser field.

Clearly, the generalized inequality calls for application to other problems \cite{benen13}. The
approach is easy to evaluate and can be used for very large systems and large
numbers of constants of motion. Thus it can prove fruitful in the intensely
studied field of integrable systems, for instance in estimating
Drude weights.
In the context of coherence in particular, various extensions of
the CSM, e.g., by magnetic fields, anisotropies, or more
intra-bath couplings suggest themselves to be investigated in the
presented manner.

\begin{acknowledgments} We gratefully acknowledge helpful discussions
with M.~Brockmann, A.~Faribault, A.~Greilich, and D.~Schuricht.  Financial support
was given by the Mercator Stiftung under Pr-2011-0003 (GSU), by the Studienstiftung des
Deutschen Volkes (DS) and by the Deutsche Forschungsgemeinschaft 
under AN 275/7-1 (FBA) and UH 90/9-1 (GSU).
\end{acknowledgments}

%\bibliography{liter10c} 

 %

%%------------------------------------------ 
\section{Supplemental Material}

\subsection{Time-Dependent Expectation Values}
\label{ss:initial}

One may wonder whether the two-time correlations $S(t)=\langle A^\dagger(t) A(0)\rangle$
reflect time-dependent measurements after the preparation of some initial state. We show
that this is the case for the simple, but important example of a spin correlation for $S=1/2$, i.e., for $A=S^z_0$.
Then we write $S^z_0=\frac{1}{2}(P_+ -P_-)$ where $P_\sigma$ projects onto the
states with $S^z_0=\sigma/2$. If $\rho$ denotes the density matrix of the total system before
any state preparation we calculate
\begin{subequations}
\begin{eqnarray}
S(t) &=& \langle S_0^z(t) S_0^z(0)\rangle
\\
&=& \frac{1}{2}\langle S_0^z(t) (P_+ -P_-)\rangle
\\
&=& \frac{1}{2}\langle S_0^z(t) P_+\rangle - \frac{1}{2}\langle S_0^z(t) P_-\rangle 
\label{eq:second}
\\
&=& \langle S_0^z(t) P_+\rangle 
\\
&=& \text{Tr}\left(S_0^z(t) P_+\rho\right)
\\
&=& \frac{1}{2}\text{Tr}\left(S_0^z(t) \rho_\text{initial} \right)
\label{eq:initial}
\\
&=& \frac{1}{2}\langle S_0^z(t)  \rangle_\text{initial}
\end{eqnarray}
\end{subequations}
where we assumed that the Hamiltonian $H$ and the density matrix $\rho$ are
invariant under total inversion $S^z\to -S^z$ so that the second term
in \eqref{eq:second} equals the first one. Finally, in \eqref{eq:initial}
we define the initial density matrix $\rho_\text{initial}:=(1/2)P_+\rho$
which results from $\rho$ by projecting it to the states with $S_0^z=1/2$
and its proper normalization. This clearly shows that in the studied case
$S(t)$ equals the time-dependent expectation value for a suitably prepared
initial state.

The above procedure can be modified to other observables. Generally, we can consider
  $\langle A^\dagger(t) D(0)\rangle$ to focus on the time-dependent expectation value
	$\langle A^\dagger(t)\rangle_D$ starting from the initial density matrix
	$\rho_\text{initial} :=D\rho$. However, do not claim that a suitable operator $D$ is
	easy to find. This route remains to be explored in future work.

\subsection{Rigorous Bound for Non-Decaying Spin Correlation}
\label{ss:bounding_elements}

For completeness, we recall the following definitions of
conserved quantities. The total angular momentum $\vec{I}$ and the combination $\vec{I}_Q$ derived from it read
\begin{subequations}
\begin{eqnarray}
\vec{I} &:=& \sum_{j=0}^N \vec{S}_j
\\
\vec{I}_Q &:=& \sum_{j=0}^N \vec{S}_j \sum_{0 \le l < p \le N}(\vec{S}_l\cdot\vec{S}_p).
\end{eqnarray}
Below we only need the corresponding $z$-components. Furthermore, we consider
\begin{equation}
H^z_l := \sum_{j=0}^N S^z_j \sum_{j=0,\ne l}^N J_j^{(l)}(\vec{S}_l\cdot\vec{S}_j)
\end{equation}
\end{subequations}
based on the constants of motion $H_l=\sum_{j=0,\ne l}^N J_j^{(l)}(\vec{S}_l\cdot\vec{S}_j)$ 
of the integrable CSM \cite{Gaudin1976,*gaudi83}
where we use the shorthand $J_j^{(l)}$ and introduce some further shorthands for future use
\begin{subequations}
\begin{eqnarray}
J_j^{(l)} &:=& (\varepsilon_l - \varepsilon_j)^{-1}
\\
S^{(l)} &:=& \sum_{j=0,\ne l}^N J_j^{(l)}
\\
Q^{(l)} &:=& \sum_{j=0,\ne l}^N \left(J_j^{(l)}\right)^2
\end{eqnarray}
\end{subequations}
where $\varepsilon_0=0$ and $\varepsilon_j=-J_j^{-1}$. Note that $J_j=J_j^{(0)}$, $J_S=S^{(0)}$, and $J_Q^2=Q^{(0)}$.

For the disordered spin system with density operator $\rho$ proportional to the identity the following
diagonal scalar products can be determined straightforwardly
\begin{subequations}
\label{eq:N-diag}
\begin{eqnarray}
(I^z|I^z) &=& (N+1)/4
\\
(I^z_Q|I^z_Q) &=& (N+1)N(7N-5)/128
\\
(H^z_l|H^z_l) &=& (2(S^{(l)})^2+3(N-1)Q^{(l)})/64.
\end{eqnarray}
\end{subequations}
We also need the  non-diagonal matrix elements
\begin{subequations}
\label{eq:N-nondiag}
\begin{eqnarray}
(I^z|I^z_Q) &=& (N+1)N/16
\\
(I^z|H^z_l) &=& S^{(l)}/8
\\
(I^z_Q|H^z_0) &=& J_S(7N-5)/64
\\
\nonumber
(H^z_l|H^z_p) &=& J_p^{(l)}(S^{(p)}-S^{(l)})/16\\
&-&3(N-3)(J_p^{(l)})^2/64\quad \text{for} \quad l\ne p. \quad
\end{eqnarray}
\end{subequations}

For the observable $S^z_0$ we obtain the vector elements
\begin{subequations}
\label{eq:a-element1}
\begin{eqnarray}
(S^z_0|I^z) &=& 1/4
\\
(S^z_0|I^z_Q) &=& N/16
\\
(S^z_0|H^z_0) &=& J_S/16
\\
(S^z_0|H^z_l) &=&  -J_l/16 \quad \text{for} \quad l>0 .
\end{eqnarray}
\end{subequations}

With these matrix and vector elements we can compute $S_\text{low}$ in \eqref{eq:gen-low-bo}
for various sets of conserved quantities. Note that $H^z_0$ is linearly dependent on
the $N$ quantities $H^z_l$ with $0 < l \le N$ due to 
\begin{equation}
\sum_{l=0}^N H^z_l =0.
\end{equation}
Similarly, $I^z_Q$ depends linearly on them due to
\begin{equation}
I^z_Q= \sum_{l=1}^N \varepsilon_l H^z_l .
\end{equation}
Hence, one may either consider $I^z$ together with the $N$
quantities $H^z_l$ with $0 < l \le N$ \emph{or} the three quantities $I^z, I^z_Q, H^z_0$.
The first choice exploits all the known conserved quantities on
the considered level of at most trilinear spin combinations.
This is what is called `all quantities' in Fig.\ 1 in the Letter.
No explicit formula can be given, but the required matrix inversion
is easily performed for up to $N=O(1000)$ spins with any computer algebra program
and up to $N\approx 10^6$ spins by any subroutine package for linear algebra.

The second choice of $I^z, I^z_Q, H^z_0$ yields $3\times3$ matrices and can
be analysed analytically. Inserting the elements in \eqref{eq:N-diag} and in \eqref{eq:N-nondiag}
and those in \eqref{eq:a-element1} into \eqref{eq:gen-low-bo} yields
\begin{equation}
\label{eq:3res}
S_\text{low}= \frac{1}{4(N+1)}\frac{(3J_Q^2+J_S^2)N(N+1)-10J_S^2}{3J_Q^2N(N+1)+2J_S^2(N-5)}.
\end{equation}
Furthermore, these three quantities are conserved for
any isotropic spin model so that we may also consider the system with
the additional bond $H=H_0+J_\text{ex}\vec{S}_1\cdot\vec{S}_N$, see Fig.\ 1.
Thus we extend the above formulae by passing from $H_0$ to $H$ and hence
from $H^z_0$ to $H^z=I^z H$. The modified scalar products are
\begin{subequations}
\begin{eqnarray}
(H^z|H^z) &=& (H^z_0|H^z_0)
\nonumber
\\
&+&(J_1+J_N)J_\text{ex}/16+ (3N-1)J_\text{ex}^2/64
\qquad
\\
(I^z|H^z) &=& (I^z|H^z_0) + J_\text{ex}/8
\\
(I^z_Q|H^z) &=& (I^z_Q|H^z_0) + (7N-5)J_\text{ex}/64
\\
(S^z_0|H^z) &=&(S^z_0|H^z_0).
\end{eqnarray}
\end{subequations}
They lead to a bound $S_\text{low}(J_\text{ex})$ as depicted in Fig.\ 1.
The explicit formula is similar to the one in \eqref{eq:3res}, but
lengthy so that we do not present it here. It can be
easily computed by computer algebra programs.

%BB
\subsection{Estimate via  Bound for the Overhauser Field}
\label{ss:bb_bounds}

Eq.\ \eqref{eq:estimate} relates the non-decaying fraction $S_\infty$
to the relative bound for the Overhauser field
\begin{equation}
\vec{B} =  \sum_{j=0}^N J_j \vec{S}_j
\end{equation}
where $J_0$ is arbitrary if the central spin has $S=1/2$. 
We stress, however, that the derivation yielding \eqref{eq:estimate}
in Ref.\ \onlinecite{merku02} only holds for the CSM
so that we do not consider extensions to finite $J_\text{ex}$ in this case.

We use the freedom to choose $J_0$
to maximize the resulting lower bound for $A=B^z$. We reuse all matrix
elements of the norm matrix $\mathbf{N}$ in \eqref{eq:N-diag} and in \eqref{eq:N-nondiag}.
Since \eqref{eq:estimate} uses the relative correlation we have to
compute 
\begin{equation}
S^{(B)}(t=0)=(B^z|B^z)=(J_Q^2+J_0^2)/4
\end{equation}
as well.
Furthermore, the vector elements of $\mathbf{a}$ must be determined anew
\begin{subequations}
\label{eq:a-element2}
\begin{eqnarray}
(B^z|I^z) &=& (J_S+J_0)/4
\\
(B^z|I^z_Q) &=& (J_S+J_0)N/16
\\
(B^z|H^z_0) &=& (J_Q^2+J_0 J_S)/16
\\
\nonumber
(B^z|H^z_l) &=&  J_l(S^{(l)}+J_l)/8\\
&-&J_l(J_S+J_0))/16 \quad \text{for} \quad l>0.
\end{eqnarray}
\end{subequations}
These elements allow us to determine the ratio $S^{(B)}_\text{low}/S^{(B)}(0)$ for the three
quantities $I^z, I^z_Q, H^z_0$ or for all quantities, i.e., $I^z$ and $H^z_l$ with $1\le l\le N$.
The ensuing  lower bounds can be optimized by varying $J_0$ in such a way that the ratios become 
maximum yielding the best bounds. The latter step is easy to perform since the non-linear
equation in $J_0$ to be solved to determine the maximum
is just a quadratic one. In this way, the triangle and square
symbols in Fig.\ 1 are computed. 

The comparison to the Bethe ansatz data for up to $N=48$ spins in Ref.\ \cite{farib13b}
yields an excellent agreement within the accuracy with which we can read off $S_\infty$ from the
numerically evaluated Bethe ansatz correlation $S(t)$. This concludes the section on the
required input of matrix and vector elements.

\end{document}